\documentclass[twocolumn]{jpsj2} 
\renewcommand{\Vec}[1]{\mbox{\boldmath$#1$}}
\title{The Origin of the Charge Ordering and Its Relevance to
Superconductivity in $\theta$-(BEDT-TTF)$_2$X: The Effect of the
Fermi Surface Nesting and the Distant Electron-Electron Interactions}
\author{
Kazuhiko Kuroki
}
\inst{
Department of Applied Physics and
Chemistry, The University of Electro-Communications,
Chofu, Tokyo 182-8585, Japan}
\abst{
The origin of the charge ordering in 
organic compounds 
$\theta$-(BEDT-TTF)$_2 X$ ($X=MM'$(SCN)$_4$, $M=$Tl,Rb,Co, $M'=$Cs,Zn) is 
studied using an extended Hubbard model.  Calculating the charge 
susceptibility within random phase approximation (RPA), we find that 
the $(3\times 3)\sim (3\times 4)$  
charge ordering observed at relatively high temperatures can be considered 
as a consequence of a 
cooperation between the Fermi surface nesting, controlled by 
the hopping integral in the $c$ direction, and the 
electron-electron interactions, where 
the distant (next nearest neighbor) interactions that 
have not been taken into account in most of the previous studies play  
an important role.
Mean field analysis at $T=0$
also supports the RPA results, and further 
shows that in the $3\times 3$ 
charge ordered state, some portions of the Fermi surface remain  
ungapped and are nested with a nesting vector close to the modulation 
wave vector of the horizontal stripe ordering 
observed at low temperatures in $X=MM'$(SCN)$_4$. 
We further study the possibility of superconductivity by taking into 
account the distant off-site repulsions and 
the band structure corresponding to $X=$I$_3$, in which 
superconductivity is experimentally observed.  We find 
that there is a close competition between  $d_{xy}$-wave-like 
singlet pairing and $p_{x+2y}$-wave-like triplet pairing  
due to a cooperation between the charge and the spin fluctuations.
The present analysis provides a possible unified understanding of the 
experimental phase diagram of the $\theta$-(BEDT-TTF)$_2 X$ family, 
ranging from a charge ordered insulator to a superconductor.
}
\kword{$\theta$-(BEDT-TTF)$_2$X, charge ordering and fluctuations,
electron-electron interactions, Fermi surface, pairing symmetry}

\begin{document}
\maketitle

\section{Introduction}

\begin{figure}
\begin{center}
\includegraphics[width=8cm,clip]{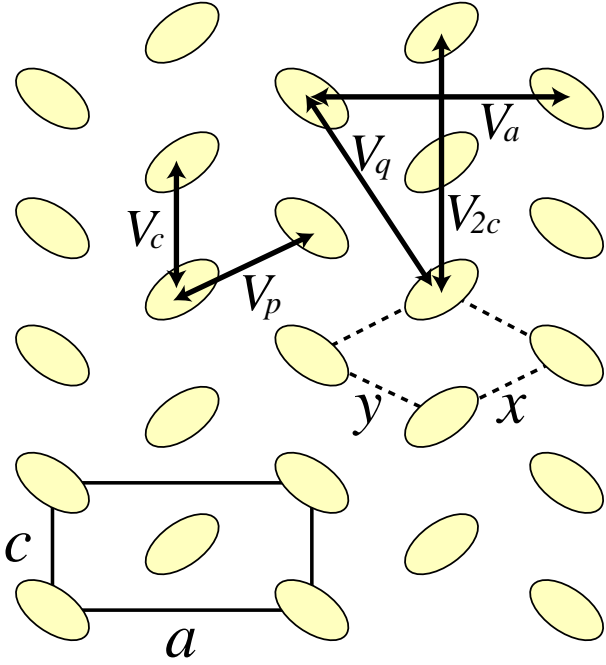}
\end{center}
\caption{The lattice structure of the cation layer of 
$\theta$-(BEDT-TTF)$_2X$. $V_p$, $V_c$ are the nearest neighbor 
interactions, while $V_a$, $V_q$, $V_{2c}$ are the next nearest neighbor 
interactions. The $a$-$c$ unit cell is the usual unit cell, while 
we can use the $x$-$y$ unit cell to unfold the BZ.}
\label{fig1}
\end{figure}

\begin{figure}
\begin{center}
\includegraphics[width=8cm,clip]{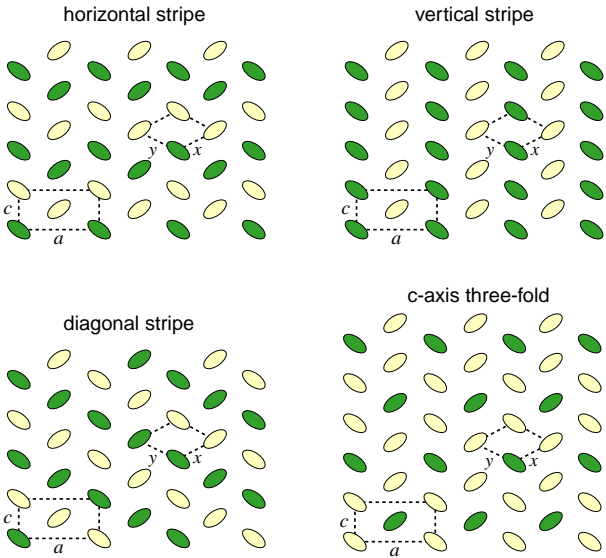}
\end{center}
\caption{(Color online) Horizontal stripe, 
vertical stripe, diagonal stripe, 
and $c$-axis three-fold charge orderings are shown. 
The green (or dark) molecules are the charge rich ones.}
\label{fig2}
\end{figure}

$\theta$-(BEDT-TTF)$_2 X$ is one of the most interesting families of 
organic compounds, ranging from charge ordered 
insulators such as $X=$RbZn(SCN)$_4$ to a superconductor $X=$I$_3$.
\cite{HMori1,HMori2}
The anion $X$ controls the angle between the BEDT-TTF molecules, 
which in turn determines the band structure.
A further fascination for this series of compounds has arisen by a 
resent observation of a giant nonlinear transport in 
$X=$Cs$M'$(SCN)$_4$ ($M'=$Co,Zn), which makes the material 
work as an organic thyristor.\cite{Inagaki,Sawano}
There, it has been pointed out that the coexistence of two 
kinds of short range charge ordering with modulation wave vectors 
$(q_a,q_b,q_c)=(\frac{2}{3},k,\frac{1}{3})$ and $(0,k,\frac{1}{2})$
(in units of the reciprocal lattice primitive vectors in $a$, $b$, $c$ 
directions; $a$ and $c$ directions are shown in Fig.\ref{fig1}) 
plays an important role in this phenomenon. 
The $(\frac{2}{3},k,\frac{1}{3})$ modulation 
corresponds to a $3\times 3$ ordering with respect to the 
$a$-$c$ unit cell shown in Fig.\ref{fig1}, while a charge modulation with 
$(0,k,\frac{1}{2})$ 
is often referred to as the horizontal stripe ordering (Fig.\ref{fig2}).
At high temperatures,
a diffuse X-ray spot is observed at $(\frac{2}{3},k,\frac{1}{3})$,
\cite{Nogami,Watanabe1999} while the $(0,k,\frac{1}{2})$ structure develops 
as temperature is lowered, and the system becomes more insulating.
When the electric field is applied, 
the $(0,k,\frac{1}{2})$ horizontal stripe ordering is degraded, 
resulting in a recovery of the metallic behavior and thus the 
nonlinear transport.
Related to this is an observation of lattice modulation with 
a wave vector $(\frac{2}{3},k,0.29\sim \frac{1}{3})$  
in $X=$CsCo(SCN)$_4$ under pressure of 10kbar,
which has been attributed to a pressure induced $2k_F$ 
charge density wave (CDW) because the modulation wave vector 
coincides with the nesting vector of the Fermi surface.\cite{Watanabe1999} 

In $X=$Rb$M'$(SCN)$_4$ ($M'=$Co,Zn) 
also, two kinds of charge ordering are involved.
At high temperatures in the metallic phase, 
diffuse X-ray  spots are observed 
at $(\frac{2}{3},k,\frac{1}{4})$ and 
$(\frac{1}{3},k,\frac{3}{4})$, which suggests the presence 
of short range $3\times 4$ charge ordering.\cite{Watanabe2004,Watanabe2005} 
Some anomalies are also observed 
in the NMR experiments in the similar temperature range.\cite{Miyagawa,Chiba}
At around 200K, 
the system undergoes a metal-insulator transition, 
accompanied by a structural phase transition into the so-called 
$\theta_d$ phase,\cite{Seo} 
in which the unit cell is doubled in the $c$ direction.
\cite{HMori2,Miyagawa,Chiba,Watanabe2004,Watanabe2005,HMori3,Tajima,Yamamoto}
In the $\theta_d$ phase, X-ray diffraction measurements have revealed that 
a long range horizontal stripe charge ordering with the modulation wave vector 
$(0,0,\frac{1}{2})$ 
takes place.\cite{HMori1,HMori2,Watanabe2004,Watanabe2005}.

Theoretically, the origin of these charge orderings has been 
an issue of great interest. Although some orderings are accompanied 
by structural phase transition or lattice modulation such as 
the horizontal stripe ordering in $X=$Rb$M'$(SCN)$_4$ and 
the $(\frac{2}{3},k,0.29\sim \frac{1}{3})$  ordering ($2k_F$ CDW)
in $X=$Cs$M'$(SCN)$_4$ under pressure, 
such a lattice modulation is not observed 
when the $(0,k,\frac{1}{2})$ (horizontal stripe) 
short range order develops in $X=$Cs$M'$(SCN)$_4$, nor 
when $3\times (3\sim 4)$ ordering occurs in $X=MM'$(SCN)$_4$ 
at ambient pressure.
Therefore, it is reasonable to assume that the electronic degrees of 
freedom takes the initiative in these charge orderings, and 
in some cases the lattice modulation occurs as a {\it consequence} of the 
electronic charge modulation.
In understanding the origin of the charge ordering 
in $\theta$-(BEDT-TTF)$_2$X from a purely electronic point of view, 
an important progress was made by Seo, who included the nearest 
neighbor off-site repulsions $V_p$ and $V_c$ 
in the model Hamiltonian (see Fig.\ref{fig1}).\cite{Seo}
There have also been studies on a model that neglects $V_c$,
\cite{Mckenzie,Hanasaki} 
 but an estimation by Mori\cite{TMori} shows that $V_p\sim V_c$ for the 
$\theta$-(BEDT-TTF) compounds.
In ref.\citen{Seo}, 
the energy of various types of charge ordering patterns,
such as horizontal, vertical, and diagonal stripes (Fig.\ref{fig2}),  
was calculated within the mean field approximation.
There, it was found, however, that the horizontal stripe charge ordering 
is not stabilized for the $\theta$-phase lattice structure,
while it is stabilized in a certain 
parameter regime in the $\theta_d$ lattice structure, which has a 
two-fold modulation in the $c$-axis direction.
Thus, within this analysis,  
the lattice modulation seems to be the cause, 
not a consequence, of the electronic charge stripe formation.
Exact diagonalization studies on these stripe orders 
have also been performed\cite{Clay,Seo2}, but 
in the mean while, an analysis by Mori in the static limit 
revealed a possibility of non-stripe charge ordering  
shown in Fig.\ref{fig2},\cite{TMori} which has a three-fold 
periodicity in the $c$-axis direction. 
We will call this the 
``$c$-axis three-fold'' charge ordering in the present paper. 
Later on, Kaneko and Ogata 
extended Seo's mean field study to take into account the possibility of 
this $c$-axis three-fold ordering,\cite{KanekoOgata}
where they found that this ordering does have 
a lower energy then the vertical, diagonal, and horizontal stripes 
when $V_p\sim V_c$. 
In these studies, a possible relation between the 
$c$-axis three-fold ordering and the diffuse X-ray spots observed at 
$(\frac{2}{3},k,\frac{1}{3}\sim\frac{1}{4})$ was suggested. 
A variational Monte Carlo study by 
Watanabe and Ogata found a subtle competition between the $c$-axis 
three-fold ordered state and the diagonal or 
vertical stripe states,\cite{WataOgata}
but quite recently, Hotta {\it et al.} 
used exact diagonalization for a spinless half-filled 
model, which can be considered as an effective model of the $\theta$-type 
compounds in the large on-site $U$ limit, and showed the 
presence of $c$-axis 
three-fold charge correlation for $V_p\sim V_c$.\cite{Hotta}

In the present study, we will further 
confirm that the $c$-axis three-fold charge correlation 
is strong when $V_p\sim V_c$. 
Moreover, we point out that while the $c$-axis three-fold charge 
ordering is uniform in the $a$-axis direction,  the diffuse X-ray spots 
are observed at $(\frac{2}{3},k,\frac{1}{3}\sim\frac{1}{4})$, 
which corresponds to orderings with a three-fold periodicity 
in the {\it a-axis direction}.  We stress here that 
the common (and thus probably the most essential) feature in the 
high temperature charge ordering is the three-fold periodicity 
in the $a$-axis direction ($q_a=\frac{2}{3}$), 
while the periodicity in the $c$-axis 
direction is sensitive to the anions and/or the pressure.
Thus, we conclude that 
neither the low temperature horizontal stripe nor the 
high temperature $3\times (3\sim 4)$ charge orderings can be understood 
within purely electronic models that consider only $V_p$ and $V_c$ 
(with $V_p\sim V_c$) as off-site repulsions.
We argue that the consideration of more distant (next nearest 
neighbor) electron-electron 
interactions is crucial, and propose that the cooperation between the 
Fermi surface nesting and the electron-electron interactions including those 
distant ones is the origin of the charge orderings.

Another aim of the present study is to 
investigate the origin of superconductivity in 
$X=$I$_3$.\cite{HKobayashi2} Since this material sits in the vicinity of the 
(nearly) charge ordered materials in the experimental phase 
diagram,\cite{HMori2}
it is interesting to investigate whether the charge 
fluctuations can give rise to the occurrence of superconductivity.
The mechanism of superconductivity in $\theta$-type compounds has been studied 
theoretically in the past,\cite{WataOgata,Merino,KobaOgata}
but the distant interactions beyond $V_p$ and $V_c$ that 
are necessary to understand the charge ordering pattern for 
$X=MM'$(SCN)$_4$ were not included there. 
Since it is natural to assume that 
the range of the off-site interaction does not change drastically 
with the change of the anion, 
we investigate the possibility of 
superconductivity in the vicinity of the charge ordering phase in a 
model that takes into account the distant interactions and the 
band structure of $X=$I$_3$.

\section{Formulation}
\subsection{Model}
\begin{figure}
\begin{center}
\includegraphics[width=8cm,clip]{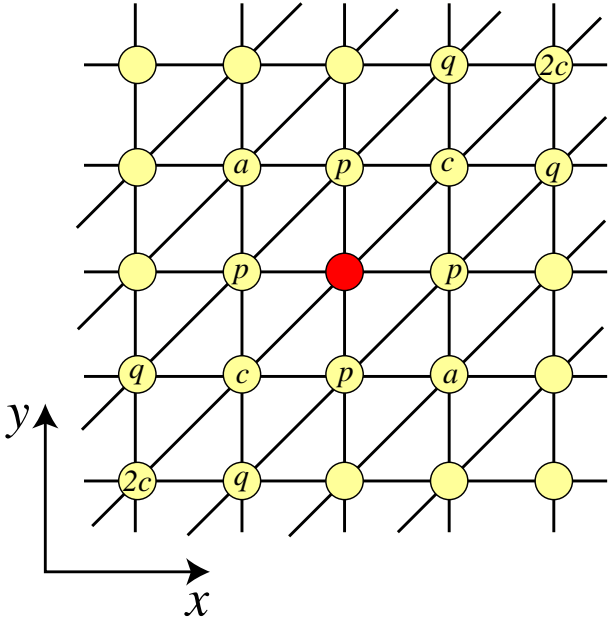}
\end{center}
\caption{(Color online) The effective lattice structure using the $x$-$y$ 
unit cell. We specify the hoppings and the interactions by 
$c$, $p$, $a$, $q$, $2c$, which specify the position relative to the 
center (red) site.}
\label{fig3}
\end{figure}

\begin{figure}
\begin{center}
\includegraphics[width=8cm,clip]{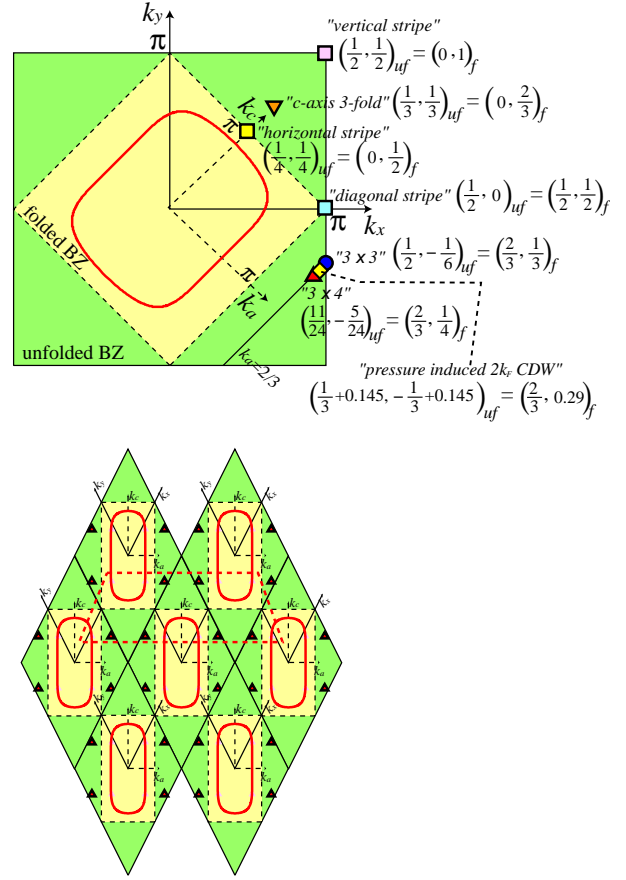}
\end{center}
\caption{(Color online) 
Upper panel: The relation between the folded and the unfolded 
BZ is shown. Modulation wave vectors corresponding to 
various orderings are shown.
The wave vectors are 
presented in units of the reciprocal lattice primitive vectors.
The red line shows the Fermi surface for $t_c=-0.3$. 
Lower panel: The BZ in the upper panel is 
squeezed in the $k_x=-k_y$ direction, and than placed repeatedly.
The trapezoid area surrounded by the red dashed line corresponds to 
the area at which the X-ray diffuse rods are observed 
in Fig.4(a) of ref.\citen{Watanabe2004}.
}
\label{fig4}
\end{figure}

In the lattice structure of $\theta$-(BEDT-TTF)$_2X$ shown in 
Fig.\ref{fig1}, the direction of the molecules alternates along 
the $a$ axis resulting in the rectangular $a$-$c$ unit cell, 
but this alternation is irrelevant as far as the 
hopping integrals in the tight binding model are concerned.
Therefore, we can take a unit cell ($x$-$y$) which is half the size of the 
usual unit cell, thereby obtaining the effective lattice structure 
shown in Fig.\ref{fig3}.
The corresponding Brillouin zone (BZ) becomes unfolded, and the 
relation between the unfolded and the folded BZ is shown 
in Fig.\ref{fig4}. Here, the wave vectors are denoted in units of the 
reciprocal lattice primitive vectors, and the subscripts $uf$ or $f$
indicate whether the reciprocal vectors are those of the unfolded or the 
folded BZ. The points 
$(\frac{2}{3},k,\frac{1}{3}\sim\frac{1}{4})_{f}$ at 
which the diffuse X-ray spots are observed in the high temperature 
regime of $X=MM'$(SCN)$_4$ fall on the diagonal line noted as 
$k_a=\frac{2}{3}$, which is a line that satisfies $k_x-k_y=\frac{2}{3}$ 
in the unfolded BZ. It can be clearly seen that these 
positions are different from the modulation wave vector of 
the $c$-axis three-fold ordering, 
$(0,\frac{2}{3})_f=(\frac{1}{3},\frac{1}{3})_{uf}$.


We consider an extended Hubbard model on the lattice shown in Fig.\ref{fig3}, 
where the Hamiltonian is given in the form 
\begin{equation}
H=\sum_{<i,j>,\sigma} 
t_{ij}c^{\dagger}_{i\sigma}c_{j\sigma}
+U\sum_{i}n_{i\uparrow}n_{i\downarrow}
+ \sum_{<i,j>}V_{ij} n_{i}n_{j},
\end{equation}
where $c^{\dagger}_{i\sigma}$ creates an electron 
with spin $\sigma = \uparrow, \downarrow$ at site $i$, 
and $n_{i\sigma}=c_{i\sigma}^\dagger c_{i\sigma}$.
The band filling, the average number of electrons per site (molecule), is 
fixed at $n=1.5$ in accord with the actual material.
In Fig.\ref{fig3}, the letters $p$, $c$, $a$, $\cdots$ denote the 
relative positions with respect to the red site in the center, 
and we use these letters to specify the range of the hopping integrals and 
the off-site repulsive interactions.
As for the kinetic energy terms, we consider hoppings  
$t_p$ and $t_c$, where $t_p(\simeq 0.1 {\rm eV})$ 
is taken as the unit of energy. 
$U$ is the on-site repulsive interaction. As for the off-site repulsions,
we consider, in addition to $V_p$ and $V_c$ (nearest neighbor interactions) 
that are usually taken 
into account, more distant interactions $V_a$, $V_q$, and $V_{2c}$ 
(next nearest neighbor interactions, see also Fig.\ref{fig1}).
Consideration of these interactions is supported by 
Mori's estimation showing that these distant interactions are 
not so small compared to $V_p$ and $V_c$.\cite{TMori} Also, there is a 
recent transport experiment which indicates that 
the repulsive interaction in 
$\theta$-(BEDT-TTF)$_2M$Zn(SCN)$_4$ ($M=$Cs,Rb) is long ranged.\cite{Yamaguchi}
In fact, the presence of such a distant interaction has also been 
pointed out theoretically for quasi-one-dimensional organic compounds 
(TMTSF)$_2$X in the context of coexisting $2k_F$ spin and $2k_F$ charge 
density waves\cite{KobayashiOgata,Suzumura,YK} and 
spin triplet pairing superconductivity.\cite{YK,KAA,KY,Fuseya,Nickel,Kuroki}
So there is a possibility that long ranged nature of the 
electron-electron interaction may be common in these organic materials.
Since the values of the electron-electron repulsions  have not so far been 
strictly estimated from, e.g., first principles calculations, 
here we vary them as parameters and investigate whether the experimentally 
observed charge orderings can be understood within a realistic 
parameter regime, where we take $V_a, V_q, V_{2c}$ to be 
smaller than $\sim V_c/2$, and $V_c\simeq V_p$ to be smaller than
$\sim U/2$.

\subsection{Random Phase Approximation}
In this subsection, we describe the random phase approximation (RPA) 
adopted in the present study.
Within RPA\cite{KobaOgata,Scalapino,TanakaOgata}, 
the spin and charge susceptibilities,  $\chi_{s}$ and 
$\chi_{c}$, respectively, are given as 
\begin{eqnarray}
\label{4}
\chi_{s}(\Vec{q})=\frac{\chi_{0}(\Vec{q})}
{1 - U\chi_{0}(\Vec{q})}
\nonumber\\
\chi_{c}(\Vec{q})=\frac{\chi_{0}(\Vec{q})}
{1 + (U + 2V(\Vec{q}) )\chi_{0}(\Vec{q})}.
\label{chargeRPA}
\end{eqnarray}
Here $\chi_{0}$ is the bare susceptibility given by 
\begin{equation}
\chi_{0}(\Vec{q})
=\frac{1}{N}\sum_{\Vec{p}} 
\frac{ f(\epsilon_{\Vec{p +q}})-f(\epsilon_{\Vec{p}}) }
{\epsilon_{\Vec{p}} -\epsilon_{\Vec{p+q}}}
\end{equation}
with
$\varepsilon_{\Vec{k}}$ being the band dispersion given as
\begin{equation}
\varepsilon_{\Vec{k}}=2t_p[\cos(k_x)+\cos(k_y)]+2t_c\cos(k_x+k_y)
\end{equation}
and 
$f(\epsilon_{\Vec{p}})=1/(\exp(\epsilon_{\Vec{p}}-\mu)/T) + 1)$ is 
the Fermi distribution function. 
When the nesting of the Fermi surface is good, 
$\chi_0(\Vec{q})$ is maximized at the nesting vector.
$V(\Vec{q})$ is the Fourier transform of the off-site repulsions, 
given as
\begin{eqnarray}
V(\Vec{q})&=&2V_p[\cos(q_x)+\cos(q_y)] \nonumber\\
&+&2V_c\cos(q_x+q_y)
+2V_a\cos(q_x-q_y)\nonumber\\
&+&2V_q[\cos(2q_x+q_y)+\cos(q_x+2q_y)]\nonumber\\
&+&2V_{2c}\cos(2q_x+2q_y)
\label{3}
\end{eqnarray}

To discuss superconductivity, 
the effective pairing interactions for the singlet and 
triplet channels due to spin and charge fluctuations are given as 
\begin{eqnarray}
\label{1}
V_{singlet}^{pair}(\Vec{q})=
U + V({\Vec q}) + \frac{3}{2}U^{2}\chi_{s}(\Vec{q})
\nonumber\\
-\frac{1}{2}(U + 2V({\Vec q}) )^{2}\chi_{c}(\Vec{q})
\label{pairsinglet}
\end{eqnarray}
\begin{eqnarray}
\label{2}
V_{triplet}^{pair}(\Vec{q})=
V({\Vec q}) - \frac{1}{2}U^{2}\chi_{s}(\Vec{q})
\nonumber\\
-\frac{1}{2}(U + 2V({\Vec q}) )^{2}\chi_{c}(\Vec{q}).
\label{pairtriplet}
\end{eqnarray}
To obtain the superconducting transition temperature 
$T_c$, we solve the linearized gap equation within the 
weak-coupling theory, 
\begin{equation}
\lambda\Delta(\Vec{k})
=-\frac{1}{N}\sum_{\Vec{k'}} V^{pair}(\Vec{k-k'})
\frac{ \rm{tanh}(\beta \epsilon_{{\Vec{k'} }}/2) }{2 \epsilon_{\Vec{k'}} }
\Delta(\Vec{k'}). 
\end{equation}
The eigenfunction $\Delta$ of this eigenvalue equation is the 
gap function. 
The transition temperature $T_c$ is determined as the temperature 
where the largest eigenvalue $\lambda$ reaches unity. 
In the actual numerical calculations, we take up to $N=256\times 256$ 
$k$-point meshes.

\subsection{Mean Field Approximation}
\label{meanfield}
In the mean field approximation, we approximate the interaction terms 
$n_{i\sigma}n_{j\sigma'}$ by 
$n_{i\sigma}\langle n_{j\sigma'}\rangle
+\langle n_{i\sigma}\rangle n_{j\sigma'}$.
We assume an electron density modulation with a modulation 
vector $(\frac{1}{2},-\frac{1}{6})_{uf}$ (two-fold periodicity in the 
$x$ direction and six-fold periodicity in the $y$ direction), 
resulting in a unit cell that contains 
six sites (see the inset of Fig.\ref{fig12} in section \ref{meanfieldres}).
This corresponds to the $3\times 3$ ordering in the original lattice 
structure because 
$(\frac{2}{3},\frac{1}{3})_{f}=(\frac{1}{2},-\frac{1}{6})_{uf}$.
Here we neglect the spin ordering for simplicity,
i.e., $\langle n_{i\uparrow}\rangle=\langle n_{i\downarrow}\rangle$, 
because (i) the ordering of the spins, 
if any, should take place in accord with the nesting vector of the 
Fermi surface, and therefore incommensurate and difficult to deal with,  
and (ii) the main aim of the mean field analysis is 
just to reinforce the RPA results and also to obtain the 
Fermi surfaces in the charge ordered state. 
We do not compare the total energy between different ordering states 
to pin down the ground state 
because (i) we do not consider the spin ordering properly and 
(ii) there could be a number of candidates for the ground state including 
incommensurate states.
We obtain the band structure in the presence of the 
$(\frac{1}{2},-\frac{1}{6})_{uf}$ ordering, and then calculate 
$\langle n_{i\sigma}\rangle$ at $T=0$,
which is substituted back into the mean field 
Hamiltonian, and the band structure is calculated again. 
These procedures are repeated until all $\langle n_{i\sigma}\rangle$ are 
self-consistently determined. We define 
\begin{equation}
\Delta n_{\rm max}=n_{\rm max\uparrow}-\langle n_\uparrow \rangle,
\label{nmax}
\end{equation}
 where 
$n_{\rm max\uparrow}$ is the largest up(=down) spin electron density among the 
six sites and $\langle n_\uparrow \rangle$ 
is the average up spin electron density, 0.75 in the present case.
In the actual numerical calculations, we take up to 
$200\times 200$ unit cells.

\section{Charge fluctuations and ordering}
\subsection{Random Phase Approximation}
\label{rpares}
\begin{figure}
\begin{center}
\includegraphics[width=8cm,clip]{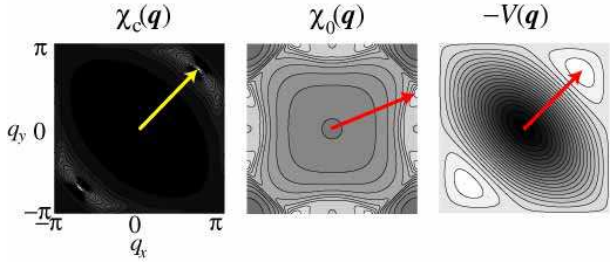}
\end{center}
\caption{(Color online) Contour plots of 
the RPA result of the charge susceptibility $\chi_c$,
the bare susceptibility $\chi_0$, and the Fourier transform of the 
off-site interactions $-V(q)$ are shown. $t_c=0$, $U=3$, $V_p=V_c=1.5$, 
$V_a=V_q=V_{2c}=0$, $T=0.05$, all in units of $t_p$.
The arrows represent the wave vector at which each function takes  
its maximum.}
\label{fig5}
\end{figure}

\begin{figure}
\begin{center}
\includegraphics[width=8cm,clip]{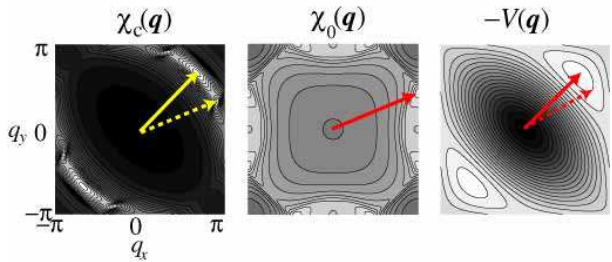}
\end{center}
\caption{(Color online) Plots similar to Fig.\ref{fig5} except 
that $V_p=1.3$.  Other parameters are the same: $t_c=0$, $U=3$, $V_c=1.5$, 
$V_a=V_q=V_{2c}=0$. 
In $\chi_c$, the dashed arrow shows the wave vector at which 
a subdominant peak exists. The dashed arrow in $-V(\Vec{q})$ indicates that 
the wave vectors at which $-V(\Vec{q})$ takes large values extend toward 
$(\frac{1}{2},0)_{uf}$ compared to the case shown in Fig.\ref{fig5}}
\label{fig6}
\end{figure}

\begin{figure}
\begin{center}
\includegraphics[width=8cm,clip]{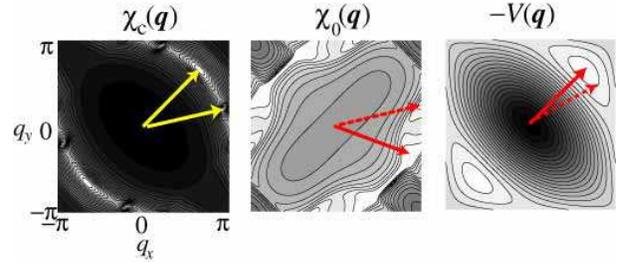}
\end{center}
\caption{(Color online) 
Plots similar to Fig.\ref{fig6} except that $t_c=-0.3$.
Other parameters are the same: $U=3$, $V_p=1.3, V_c=1.5$, 
$V_a=V_q=V_{2c}=0$. The dashed arrow in $\chi_0(\Vec{q})$ indicates that 
the wave vectors at which $\chi_0(\Vec{q})$ takes large values extend toward  
the end point of this arrow.}
\label{fig7}
\end{figure}

First we consider the case when $t_c=0$ and $V_a=V_q=V_{2c}=0$.
%
In Fig.\ref{fig5}, we show the RPA charge susceptibility 
for $U=3$ and $V_c=1.5$, 
In the case of $V_p=1.5$, there is a peak 
at $(\frac{1}{3},\frac{1}{3})_{uf}$, which corresponds to 
$(0,\frac{2}{3})_f$ in the original BZ, 
namely, the modulation wave vector of the 
$c$-axis three-fold ordering. This is due to a peak 
in $-V(\Vec{q})$ at $\Vec{q}=(\frac{1}{3},\frac{1}{3})_{uf}$ as shown 
in Fig.\ref{fig5}. Namely, although $\chi_0(\Vec{q})$ is 
maximized around $(0,\pm\frac{1}{6})_{uf}$ and 
$(\pm\frac{1}{6},0)_{uf}$, the effect of 
 $-V(\Vec{q})$ dominates in the denominator of the RPA formula of the 
charge susceptibility eq.(\ref{chargeRPA}),
thereby minimizing the denominator and maximizing $\chi_c(\Vec{q})$ at 
$(\frac{1}{3},\frac{1}{3})_{uf}$.
For $V_p$ slightly smaller than $V_c$, which is more
realistic for $\theta$-(BEDT-TTF) compounds\cite{TMori}, the peak at 
$(\frac{1}{3},\frac{1}{3})_{uf}$ still remains (Fig.\ref{fig6}), 
but there also 
appears a subdominant peak around a wave vector somewhat close to 
$(\frac{1}{2},0)_{uf}$, i.e., the modulation vector of the 
diagonal stripe. The origin of this second 
peak becomes more clear if we further introduce a finite $t_c=-0.3$, 
which roughly corresponds to the case of $X=$Rb$M'$(SCN)$_4$ ($M'$=Zn, Co) or 
$X=$CsCo(SCN)$_4$ under pressure.\cite{Watanabe1999}. In this case, 
the nesting of the Fermi surface (see Fig.\ref{fig4}) becomes good
around $(\frac{2}{3},\frac{1}{4})_{f}=(\frac{11}{24},-\frac{5}{24})_{uf}$, 
as can be 
seen from the bare susceptibility $\chi_0$ shown in Fig.\ref{fig7}. 
Now the peak near 
$(\frac{1}{2},0)_{uf}$ has about the same height as that at 
$(\frac{1}{3},\frac{1}{3})_{uf}$, 
which is due to 
a combination of $\chi_0(\Vec{q})$ that has large values extending 
from the nesting position toward $(\frac{1}{2},0)_{uf}$,  and 
$-V(\Vec{q})$  in which
large values extend to wave vectors 
closer to $(\frac{1}{2},0)_{uf}$ when $V_p<V_c$.

Thus, in all of the cases studied above, the peak position of the 
charge susceptibility disagree with the experimental observations.
We have looked into various other cases with $V_c\sim V_p$ and 
$V_a=V_q=V_{2c}=0$, where we always 
found that the charge susceptibility does not have a peak at 
$(\frac{1}{2},-\frac{1}{6})_{uf}\sim (\frac{11}{24},-\frac{5}{24})_{uf} 
=(\frac{2}{3},\frac{1}{3}\sim\frac{1}{4})_{f}$, 
corresponding to $3\times (3\sim 4)$ modulation,  
nor at $(\frac{1}{4},\frac{1}{4})_{uf}=(0,\frac{1}{2})_{f}$,  
which corresponds to the horizontal stripe modulation.

\begin{figure}
\begin{center}
\includegraphics[width=8cm,clip]{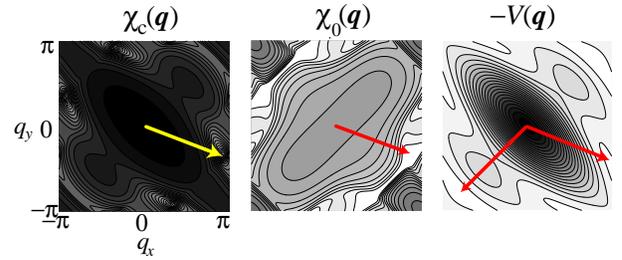}
\end{center}
\caption{(Color online) Plots similar to Fig.\ref{fig7} except that $V_a=0.3$, 
$V_q=0.5$, $V_{2c}=0.7$. 
Other parameters are the same: $t_c=-0.3$, $U=3$,  $V_p=1.3, V_c=1.5$, 
$T=0.05$.}
\label{fig8}
\end{figure}

\begin{figure}
\begin{center}
\includegraphics[width=8cm,clip]{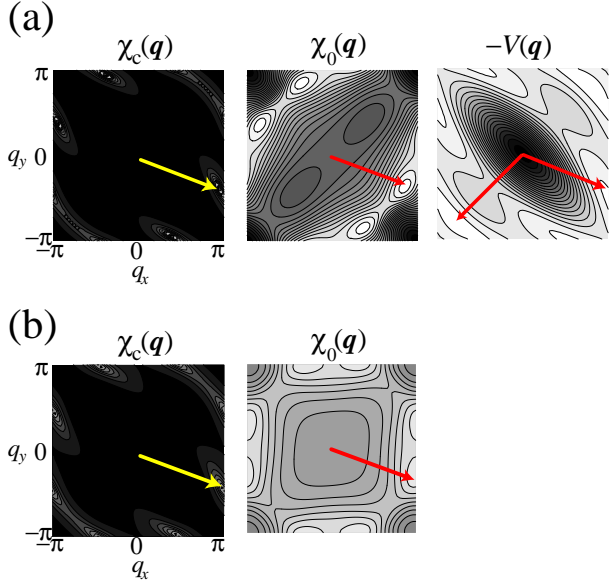}
\end{center}
\caption{(Color online) (a) Plots similar to Fig.\ref{fig8} with  
$t_c=-0.3$, $U=4$, $V_p=1.8$, $V_c=2.0$, $V_a=0.4$, $V_q=0.7$, $V_{2c}=1.1$, 
$T=0.25$. (b) Plots similar to (a) except that $t_c=-0.05$. 
$V(\Vec{q})$ is the same with that shown in (a).}
\label{fig9}
\end{figure}


Thus the model that considers only $U$, $V_p$, and $V_c$ as 
electron-electron repulsions fails 
to explain the experimentally observed charge orderings.
Since we believe that the electronic degrees of freedom is 
responsible for these orderings,  we now turn on 
the distant interactions $V_a$, $V_q$, $V_{2c}$. 
In Figs.\ref{fig8} and \ref{fig9}(a), 
we show the charge susceptibility for $U=3$ and 
$U=4$ with $t_c=-0.3$, 
maintaining the ratios of the off-site repulsions to $U$ roughly 
the same. In both cases we find a peak near the 
positions at which the diffuse X-ray spots are 
observed in the metallic phase of $X=MM'$(SCN)$_4$. This is due to the 
fact that in these cases, $-V(\Vec{q})$ is now broadly 
maximized in a region that includes the nesting vector position, as shown in 
Figs.\ref{fig8} and \ref{fig9}(a). 
Therefore, the denominator in eq.(\ref{chargeRPA}) is now 
minimized at a position close to the nesting vector, resulting in 
the charge susceptibility peak position that nearly 
coincides with the diffuse X-ray spot positions. 
\begin{figure}
\begin{center}
\includegraphics[width=8cm,clip]{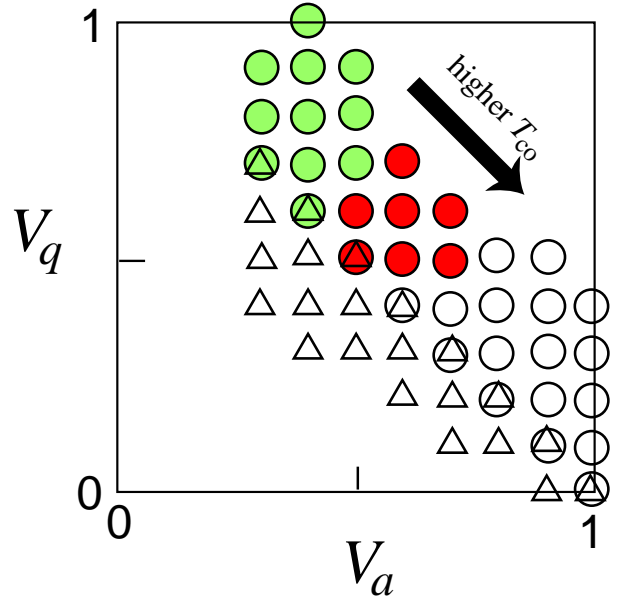}
\end{center}
\caption{(Color online) 
Marked by circles $(V_{2c}=1.1)$ or by triangles $(V_{2c}=0.7)$ 
are the values of $(V_a,V_q)$ for which the 
charge susceptibility peak exists near the 
diffuse X-ray spot positions in the metallic phase (see text)
for $t_c=-0.3$, $U=4$, $V_p=1.8$, and $V_c=2.0$. The red (or dark) 
[green (or light)] circles are the values of $(V_a,V_q)$ for 
which triplet [singlet] 
pairing superconductivity dominates for $t_c=+0.2$ (see 
section{\ref{rpares}}). 
For $(V_a,V_q)$ with uncolored (open) circles, charge ordering occurs 
above $T=0.2$ for $t_c=+0.2$.}
\label{fig10}
\end{figure}

\begin{figure}
\begin{center}
\includegraphics[width=8cm,clip]{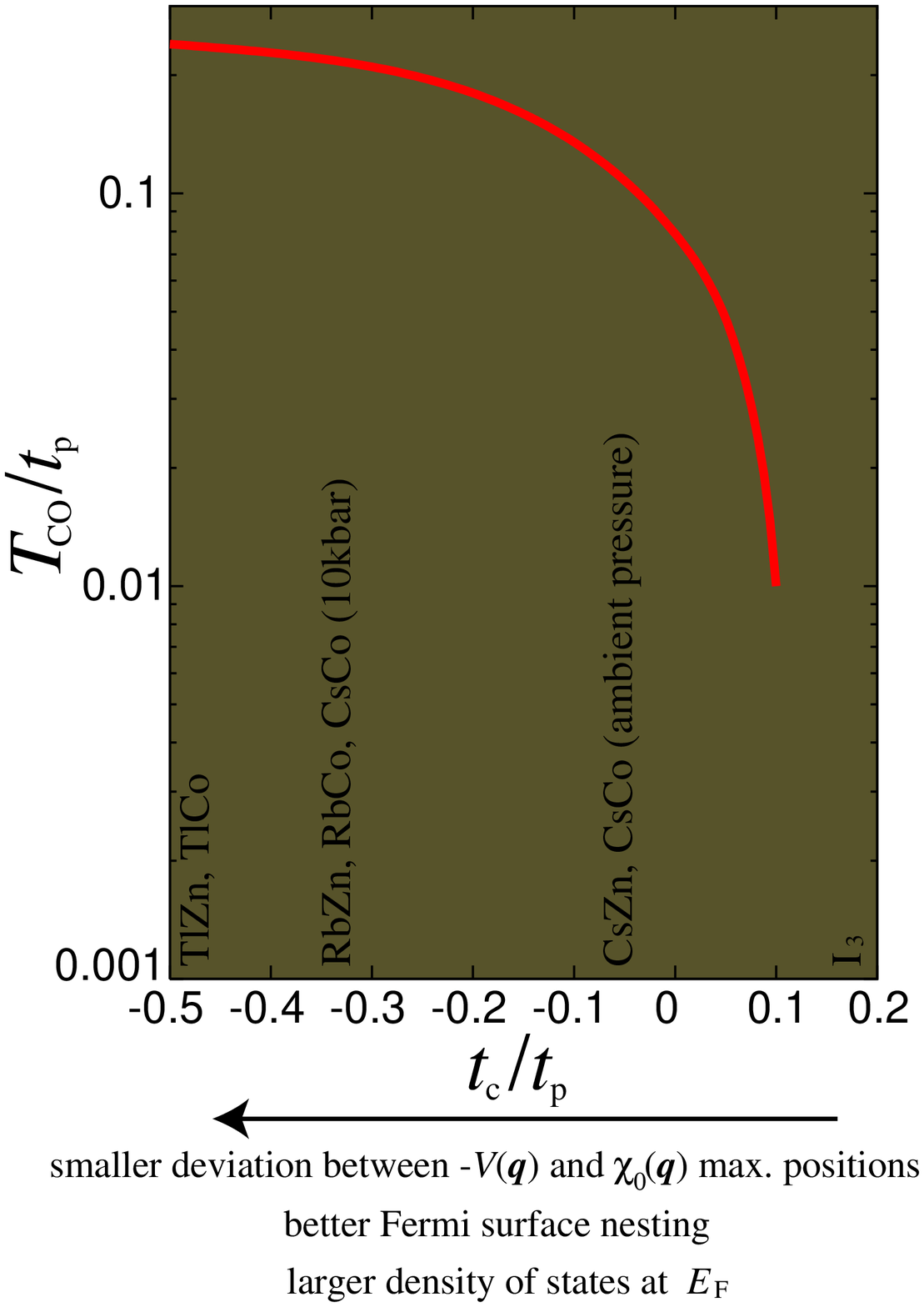}
\end{center}
\caption{(Color online) 
$T_{\rm co}$ is plotted as a function of $t_c$ for 
$U=4$, $V_p=1.8$, $V_c=2.0$, $V_a=0.4$, $V_q=0.7$, and $V_{2c}=1.1$.
$t_p\simeq 0.1eV\sim 1000$K is taken as the unit of temperature.
The values of $t_c/t_p$ for $X=MM'$(SCN)$_4$ has been taken from 
table II in ref.\citen{Watanabe1999}. As for $X=$I$_3$, 
the value of $t_p$ is taken from ref.\citen{HMori2} while 
that of $t_c$ is from ref.\citen{HMoriPv}.}
\label{fig11}
\end{figure}

In Fig.\ref{fig9}(b), we show the result for $t_c=-0.05$ with 
other parameters being the same with (a). This choice of $t_c$ 
corresponds to $X=$Cs$M'$(SCN)$_4$ at ambient pressure.\cite{Watanabe1999}
In this case, the peak position of the charge susceptibility 
moves toward the unfolded BZ edge, i.e., closer to 
$(\frac{1}{2},-\frac{1}{6})_{uf}=(\frac{2}{3},\frac{1}{3})_f$, 
as compared to the case with $t_c=-0.3$. 
This is because the nesting vector 
of the Fermi surface is closer to $(\frac{2}{3},\frac{1}{3})_f$
as can be seen in the figure of $\chi_0$. This tendency 
is again consistent with the experimental fact that 
diffused X-ray spots at high temperatures 
are observed at $q_c=\frac{1}{3}$  
in $X=$Cs$M'$(SCN)$_4$ at ambient pressure, while 
they are at $\frac{1}{4}\leq q_c \leq \frac{1}{3}$ in 
$X=$Cs$M'$(SCN)$_4$ under  pressure and in $X=$Rb$M'$(SCN)$_4$. 

Here we have presented the charge susceptibility 
for certain choices of interaction values, but 
we have also investigated cases with other sets of parameter values 
and found similar peak positions of the charge susceptibility 
within a certain parameter regime. In Fig.\ref{fig10}, 
we show the values of the distant interactions for which 
the charge susceptibility has a peak close to the 
the diffuse X-ray spot positions, 
where ``close'' means $(q_x,q_y)_{uf}$ within the range of 
$0.45\leq q_x\leq \frac{1}{2}$ and $-\frac{1}{4}\leq q_y \leq -\frac{1}{6}$. 
As seen from this figure, we may say that good agreement with the 
experimental observation,  as far as the 
peak {\it position} of the charge susceptibility is concerned, 
is obtained when $V_a+V_q$ is 
roughly constant for a given $V_{2c}$. 
The peak {\it height} (the strength of the charge fluctuations) 
on the other hand becomes higher for larger $V_a$ and smaller $V_q$, 
which is because $-V(\Vec{q})$ around 
$\Vec{q}=(\frac{1}{2},-\frac{1}{6})_{uf}
\sim (\frac{11}{24},-\frac{5}{24})_{uf}$ is larger in this 
case, so that the cooperation between the off-site repulsions 
and the Fermi surface nesting becomes more effective.

Thus, in our present view, the $3\times (3\sim 4)$
charge ordering found in the 
high temperature metallic phase 
of $X=MM'$(SCN)$_4$ occurs due to a cooperation between the 
Fermi surface nesting and the effect of the off-site interactions 
including not only $V_c$ and $V_p$ but also $V_a$, $V_q$, $V_{2c}$.
Note that the term ``charge ordering'' is usually used as 
a real space ordering of charges that is caused by the repulsive 
electron-electron interactions (regardless of the shape of the Fermi surface), 
while a ``charge density wave'' is caused mainly by the nesting of the Fermi 
surface. 
Since the Fermi surface nesting and the electron repulsion cooperate 
in the present view, the charge ordering in the metallic phase of 
$X=MM'$(SCN)$_4$ can be considered as a mixture of these two concepts.
In this sense, the $3\times (3\sim 4)$ charge ordering is 
conceptually different from the $c$-axis three-fold charge ordering 
in that the latter has nothing to do with the Fermi surface nesting.
We believe that chances of the diffuse X-ray spots {\it accidentally} 
coinciding with the nesting vector are very slim, 
and that the Fermi surface nesting must have something to do with 
the $3\times (3\sim 4)$ charge ordering, while on the other hand, 
we do believe that the electronic degrees of freedom, not the lattice, 
is the origin of this ordering.

Now, to further show that the value of $t_c$ is important for the 
charge fluctuations and ordering, we show in Fig.\ref{fig11} 
the charge ordering temperature $T_{\rm co}$, defined as the temperature 
at which the RPA charge susceptibility diverges at a certain $\Vec{q}$, 
as a function of $t_c$. $T_{\rm co}$ decreases with the increase of 
$t_c$ (from $t_c<0$ to $t_c>0$), 
and sharply drops for $t_c>0$. In fact, the increase of $t_c$ induces 
several effects. First, the nesting of the Fermi surface is 
degraded (compare Fermi surfaces in Fig.\ref{fig4} $(t_c<0)$ 
and Fig.\ref{fig15} $(t_c>0)$), 
secondly, the nesting vector deviates from the region where 
$-V(\Vec{q})$ is maximized, and thirdly, 
the density of states near the Fermi level becomes smaller.\cite{Hottacomment}
These factors cooperate to work destructively against charge ordering.

This $t_c$ dependence of $T_{\rm co}$ 
is reminiscent of the phase diagram of the $\theta$-(BEDT-TTF) family 
obtained by Mori {\it et al.}\cite{HMori2}, where charge ordering 
takes place 
for compounds having large negative $t_c$ such as $X=$Rb$M'$(SCN)$_4$, 
while it does not for compounds having small 
or positive $t_c$ like $X=$I$_3$.
However, note that  in Fig.\ref{fig11},  
the charge ordering takes place at $T_{\rm co}$ 
with a modulation wave vector close to 
the nesting vector of the Fermi surface 
(around $(\frac{2}{3},\frac{1}{3}\sim\frac{1}{4})_f$ 
for $t_c<0$), 
while the charge ordering temperature 
in the experimental phase diagram is those for the 
$(0,k,\frac{1}{2})_f$ ordering, i.e., the horizontal stripe state.
In fact, although we have succeeded in 
understanding the high temperature $3\times (3\sim 4)$ 
ordering, the horizontal $(0,k,\frac{1}{2})_f$ 
ordering remains to be explained.
Namely, there is no trace of enhanced charge fluctuations around 
$(\frac{1}{4},\frac{1}{4})_{uf}=(0,\frac{1}{2})_{f}$ in the 
charge susceptibility shown in Figs.\ref{fig8} and \ref{fig9}.
We will come back to this point in the latter part of the next section.

\subsection{Mean Field Approximation}
\label{meanfieldres}
\begin{figure}
\begin{center}
\includegraphics[width=8cm,clip]{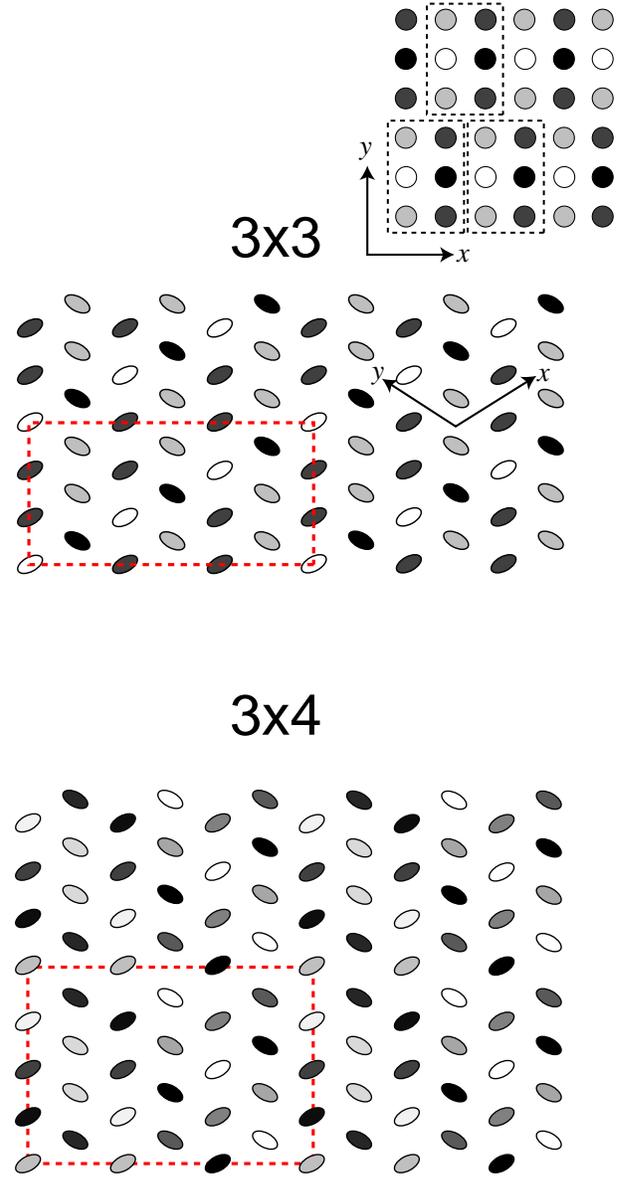}
\end{center}
\caption{(Color online) 
$3\times 3$ and $3\times 4$ charge ordering patterns are shown.
The dashed red rectangles represent the unit cell.In the upper inset of 
the $3\times 3$ ordering, the corresponding pattern on the effective 
$x$-$y$ square lattice, for which the actual mean field calculation was 
done, is displayed along with the unit cell that contains 6 sites.}
\label{fig12}
\end{figure}

We now move on to the mean field analysis at $T=0$. 
As mentioned in section \ref{meanfield}, we assume $3\times 3$ 
ordering shown in Fig.\ref{fig12}. The actual calculation is done 
for the ordering shown in the inset.
We take values of the interactions to be slightly larger than 
those adopted in the RPA calculation, $U=5$, $V_p=2.2$, $V_c=2.5$, 
$(V_a,V_q,V_{2c})=\alpha (0.5,0.8,1.2)$, thereby 
roughly maintaining the ratios of the off-site repulsions to $U$ when 
$\alpha=1$.\cite{comment} 
$\alpha$ is a parameter introduced so as to investigate the 
effect of the distant interactions in a continuous manner.

\begin{figure}
\begin{center}
\includegraphics[width=8cm,clip]{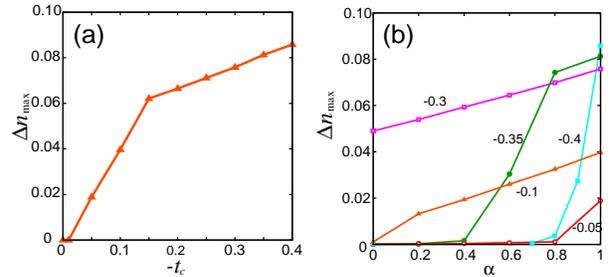}
\end{center}
\caption{(Color online) 
(a) $\Delta n_{max}$ obtained in the mean field calculation plotted 
as a function of $t_c$. (b) $\Delta n_{max}$ for various choices of $t_c$ 
plotted as functions of 
the parameter $\alpha$ that controls the distant (next nearest neighbor) 
part of the off-site repulsions.}
\label{fig13}
\end{figure}

\begin{figure}
\begin{center}
\includegraphics[width=8cm,clip]{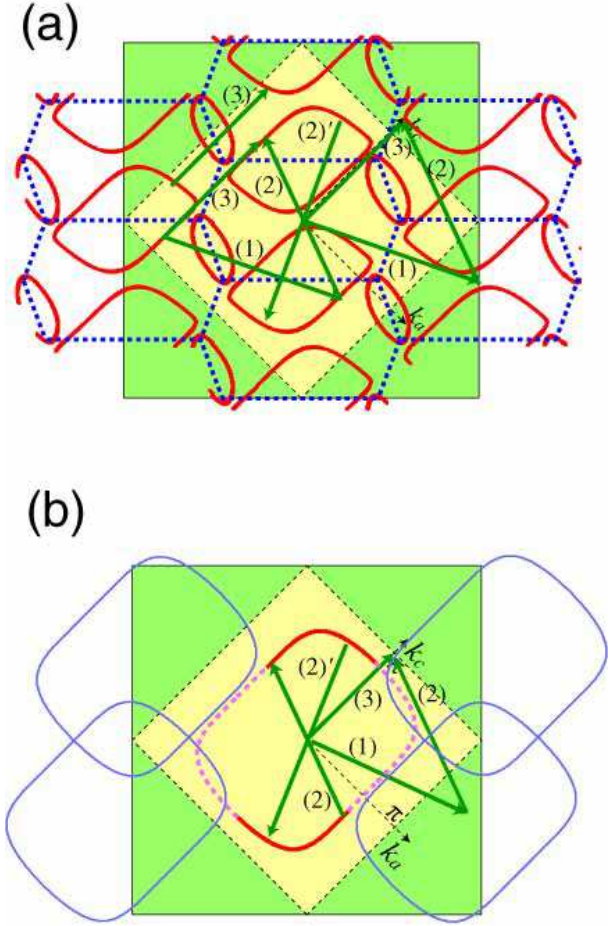}
\end{center}
\caption{(Color online) 
(a) Fermi surface of the $3\times 3$ charge ordered state 
obtained by the mean field approximation. 
$U=5.0$, $V_p=2.2$, $V_c=2.5$, $V_a=0.5$, $V_q=0.8$, $V_{2c}=1.2$, 
and $t_c=-0.3$. 
The original 
folded and unfolded BZ are also shown. 
The dashed lines represent the boundaries of the new BZ. 
See text for the explanation of the vectors (1)$\sim$ (3).
(b) The expected Fermi surface 
for the $3\times 4$ charge ordered state (see text).}
\label{fig14}
\end{figure}

In Fig.\ref{fig13}(a), 
we show $\Delta n_{\rm max}$ defined by eq.(\ref{nmax})
as a function of $t_c$ for $\alpha=1$. Here again, we find a similar 
tendency as found in the RPA calculation, namely, large 
negative $t_c$ induces the $3\times 3$ charge ordering. 
In Fig.\ref{fig13}(b), we plot 
$\Delta n_{\rm max}$ as functions of $\alpha$ for various values of $t_c$.
Here also, we find a tendency that the distant interactions induce the 
$3\times 3$ charge ordering. For $t_c=-0.3$, we find a finite 
$\Delta n_{\rm max}$ solution even for $\alpha=0$, which is probably 
because the nesting vector of the Fermi surface 
best matches the $3\times 3$ ordering at this particular $t_c$, 
but we believe that the $3\times 3$ charge ordering state is not the 
ground state at least when $\alpha=0$ considering our RPA results. 
(Note that we do not compare the total energy between different states 
as mentioned in section\ref{meanfield}. )

Now we turn to the Fermi surface in the charge ordered state.
In Fig.\ref{fig14}(a), 
we show the Fermi surface in the $3\times 3$ charge 
ordered state for $t_c=-0.3$ and $\alpha=1$. 
Note that the new BZ and the Fermi surface here are 
``unfolded'' in the sense that the unit cell in the 
inset of Fig.\ref{fig12} neglects the alternation of the molecules in 
the $a$-direction. (Therefore, the area of the 
new BZ is not 1/9 of the folded BZ but 1/6 of the unfolded BZ.)
We find that some portions of the Fermi surface remain  
ungapped, so that the system should be metallic, which 
is consistent with the experiments, where the system remains metallic 
when  diffuse X-ray 
spots are observed at $(\frac{2}{3},k,\frac{1}{3}\sim \frac{1}{4})_f$.
What is even more interesting is that there is a 
nesting of the remaining Fermi surface with a nesting vector (3), 
which is close to $(\frac{1}{4},\frac{1}{4})_{uf}=(0,\frac{1}{2})_{f}$, 
i.e., the modulation vector of the horizontal stripe ordering.
This ``new nesting'' originates from the fact that there are 
actually two different nesting vectors in the original system by symmetry, 
$(+\frac{2}{3},\frac{1}{3}\sim \frac{1}{4})_f$ and 
$(-\frac{2}{3},\frac{1}{3}\sim \frac{1}{4})_f$.
The former nesting is used to form the present $3\times 3$ 
ordering, while the latter nesting (vector (2) and $(2)'$, where 
$(2)=(2)'+(0,1)_{uf}$) 
remains in the ordered state,\cite{comment4}  
although modified to some extent from the original nesting vector 
because of the deformation of the Fermi surface. 
In Fig.\ref{fig14}(a), 
vector (1)=$(+\frac{2}{3},\frac{1}{3})_f$ is a reciprocal lattice vector 
of the $3\times 3$ ordered state at which a finite Fourier component of the 
charge density exists, so that the remaining portions of the 
original Fermi surface translated by vector (1) can 
interact with one another in the ordered state. 
Since $(2)+(1)=(3)$ holds, 
the original nesting at vector (2) results 
in a ``new nesting'' of the Fermi surface at vector (3) in the ordered state, 
which is close to 
$(0,\frac{1}{2})_f$ especially when the $c$ component of 
vector (2) (modified due to ordering) is relatively small.
Since we know from the RPA results that 
the charge fluctuations are enhanced 
in the non-ordered state around vector (2) 
due to the electron-electron interactions,
we may expect that the charge fluctuations are enhanced around the 
new nesting vector (3) in the $3\times 3$ ordered state 
due to the electron-electron interactions,  
although this remains to be confirmed in a future study. \cite{comment3} 


Although our mean field analysis is restricted to 
the $3\times 3$ 
charge ordering, we can predict what the Fermi surface should look like in the 
case of $3\times 4$ ordering from the analogy to the $3\times 3$ case.
In Fig.\ref{fig14}(b), we show the expected Fermi surface 
when the $3\times 4$ ordering
occurs, where the dashed pink lines are the portions that are expected to 
vanish due to the $(+\frac{2}{3},\frac{1}{4})_f$ 
nesting of the original Fermi surface. 
The remaining Fermi surface in the $3\times 4$ ordered state 
is expected to be nested to some extent with the nesting vector 
close to $(-\frac{2}{3},\frac{1}{4})_f$ (vectors $(2)$ and $(2)'$, 
where $(2)=(2)'+(0,1)_{uf}$).  In Fig.\ref{fig14}(b), 
$(2)+(1)=(3)$ holds, where 
vector $(1)=(+\frac{2}{3},\frac{1}{4})_{f}$ is a reciprocal lattice vector 
of the $3\times 4$ ordered state at which a finite Fourier component of the 
charge density exists. The resulting 
``new nesting vector'' (3) is precisely $(0,\frac{1}{2})_f$ when 
vector (2) is $(-\frac{2}{3},\frac{1}{4})_f$.
This precise coincidence between the new nesting vector and the 
horizontal stripe modulation vector comes from the fact that 
$q_c=\frac{1}{4}$ of both the $3\times 4$ modulation wave vector (1) and the 
nesting vector (2)
is exactly half of $q_c=\frac{1}{2}$ 
of the horizontal stripe modulation vector.

Now let us recall that within the RPA approach, we could not find 
any trace of $(0,\frac{1}{2})_{f}$ charge fluctuations.
We propose here a possibility that the new $\sim (0,\frac{1}{2})_{f}$ 
Fermi surface nesting in the $3\times (3\sim 4)$ 
charge ordered metallic state  
triggers, or at least plays some role in, the occurrence of the 
horizontal (short or long ranged) charge ordering in 
$X=MM'$(SCN)$_4$ at low temperatures.
In this view, the two charge orderings, 
the horizontal stripe 
and $3\times (3\sim 4)$, are not 
independent competing  orderings, but the former is ``based'' on the 
latter, and therefore 
we may call the $3\times (3\sim 4)$ ordering 
the ``first stage ordering'',
and the horizontal stripe ordering the ``second stage''.
In this sense, the horizontal stripe should be more ``fragile'' 
as compared to $3\times (3\sim 4)$, 
which explains why it is the horizontal stripe ordering that is 
destroyed by applying the electric field in $X=$Cs$M'$(SCN)$_4$.\cite{Sawano}
Also, when the second stage ordering takes place, most of the 
remaining ungapped Fermi surface in the first stage ordering 
should disappear, which also 
explains the (nearly) insulating behavior when the horizontal stripe 
ordering sets in. 

Let us now go back to Fig.\ref{fig11} obtained within RPA. 
In the present terminology, $T_{\rm co}$ is the temperature where the 
first stage charge ordering takes place. 
If the second stage charge ordering is indeed 
based on the first stage ordering, then we may consider that this 
$t_c$ dependence of $T_{\rm co}$ should also roughly represent the 
$t_c$ dependence of the temperature of the second stage ordering 
at $(0,\frac{1}{2})_{f}$. In this sense, we may say that 
Fig.\ref{fig11} is closely related to the experimental 
phase diagram by Mori et al.\cite{HMori2} 
This view is also consistent with an experimental 
analysis under uniaxial pressure, which concludes that 
the value of $t_c$ is the key that dominates the occurrence of the 
horizontal charge ordering.\cite{Kagoshima2}

\section{Possibility of Superconductivity: an RPA analysis}

Having found that the distant off-site interactions are necessary 
in order to understand the charge orderings and fluctuations of 
$\theta$-(BEDT-TTF) 
compounds with $t_c<0$, we now investigate the possibility of 
superconductivity induced by such charge fluctuations.
Since superconductivity is observed in $X=$I$_3$, which has 
a positive $t_c$,\cite{HMoriPv,HKobayashi,Tamura2} 
we concentrate on the case of $t_c>0$.
In fact, the value $t_c/t_p\simeq 0.19$  estimated by Mori for $X=$I$_3$,
\cite{HMoriPv} falls  close to a point at which $T_{\rm co}$ falls 
rapidly in Fig.\ref{fig11}, so that 
a charge fluctuation mediated superconductivity is possible.
Here we mainly focus on two sets of off-site interaction 
values that give the 
correct charge susceptibility peak positions at $t_c=-0.3$, 
i.e., $U=4,$ $V_p=1.8$, $V_c=2$ with $V_a=0.4$, $V_q=0.7,$ $V_{2c}=1.1$ 
or with $V_a=0.6$, $V_q=0.6,$ $V_{2c}=1.1$ (see Fig.\ref{fig10}).
Using the RPA+gap equation approach, we search for superconductivity 
within the temperature range of $T>0.002t_p$ 
under the condition that the denominator in the charge 
susceptibility formula (eq.(\ref{chargeRPA})) is larger than 0.02.

\begin{figure}
\begin{center}
\includegraphics[width=8cm,clip]{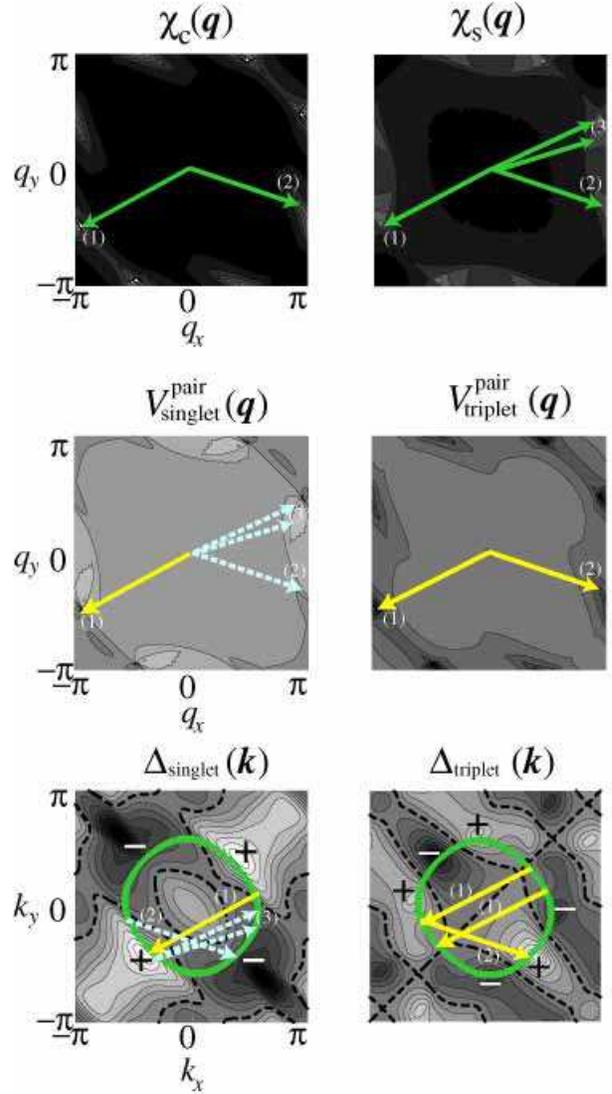}
\end{center}
\caption{(Color online)  Contour plots of the 
spin and charge susceptibilities, singlet and triplet 
pairing interactions, and the singlet ($d_{xy}$-wave like) 
and triplet ($p_{x+2y}$-wave like) gap functions for 
$U=4.0$, $V_p=1.8$, $V_c=2.0$, $V_a=0.4$, $V_q=0.7$, $V_{2c}=1.1$, 
$t_c=0.14$, and $T=0.01$. The gap functions are those that give 
the largest eigenvalue $\lambda$ for the singlet and triplet pairing channels.
The Fermi surface for $t_c=0.14$  (green solid line) 
is superposed in the figures of the 
gap function. 
The dashed lines in the gap functions represent the nodal lines.
Spin and charge fluctuations around 
wave vectors $(1)\sim (3)$ play important roles in determining the 
pairing symmetry (see text).  The yellow solid (light blue dashed) 
arrows indicate positive (negative) values of the pairing interactions.}
\label{fig15}
\end{figure}

\begin{figure}
\begin{center}
\includegraphics[width=8cm,clip]{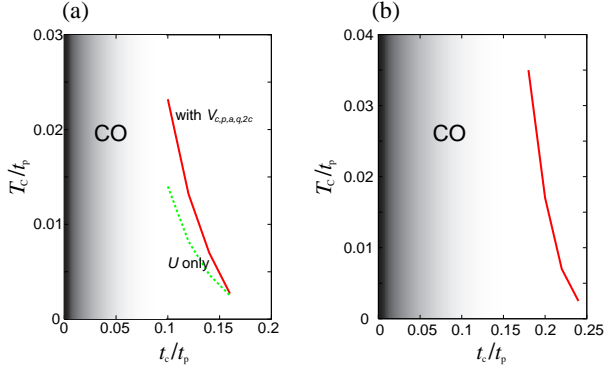}
\end{center}
\caption{(Color online) (a) Red solid line : $T_c$ for the singlet 
$d_{xy}$-wave pairing as a function of $t_c$ with 
$U=4.0$, $V_p=1.8$, $V_c=2.0$, $V_a=0.4$, $V_q=0.7$, and $V_{2c}=1.1$.
Green dashed line: Same as in the red solid line except 
all the off-site repulsions are turned off. (b) $T_c$ for the 
triplet $p_{x+2y}$-wave pairing as a function of 
$t_c$ with $U=4.0$, $V_p=1.8$, $V_c=2.0$, 
$V_a=0.6$, $V_q=0.6$, and $V_{2c}=1.1$.}
\label{fig16}
\end{figure}

\begin{figure}
\begin{center}
\includegraphics[width=8cm,clip]{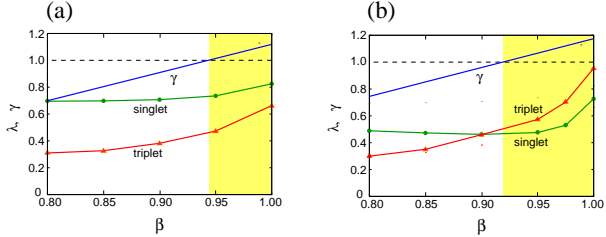}
\end{center}
\caption{(Color online) 
(a) The eigenvalues of the gap equation for the 
spin singlet and triplet pairings, along with the ratio 
$\gamma$ that measures the relative strength ratio between the 
spin and charge fluctuations (see text), are plotted as 
functions of $\beta$, where the distant interactions 
are taken as $(V_p,V_c,V_a,V_q,V_{2c})=\beta(1.8,2.0,0.4,0.7,1.1)$.
$U=4.0$, $t_c=0.16$, and $T=0.02$ are taken. The yellow hatched area 
is the regime where charge fluctuations dominate over 
spin fluctuations.
(b) Same with (a) except 
$(V_p,V_c,V_a,V_q,V_{2c})=\beta(1.8,2.0,0.6,0.6,1.1)$ and 
$t_c=0.2$.}
\label{fig17}
\end{figure}

\begin{figure}
\begin{center}
\includegraphics[width=8cm,clip]{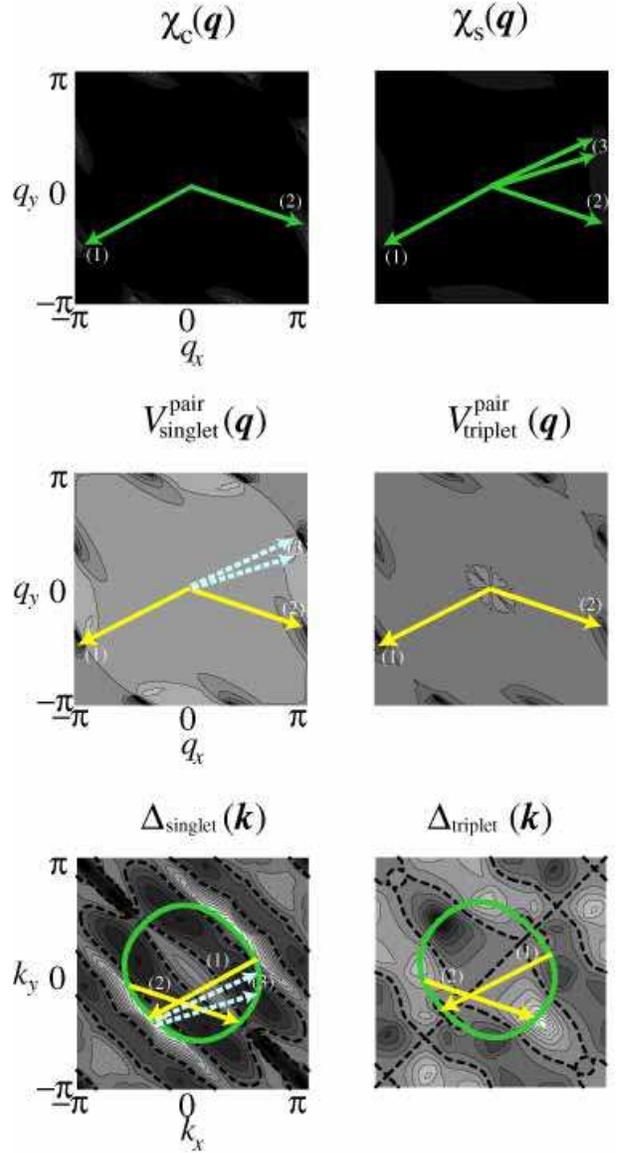}
\end{center}
\caption{(Color online) Plots similar to Fig.\ref{fig15} except that 
$V_a=0.6$, $V_q=0.6$, $V_{2c}=1.1$, and $t_c=0.2$.  }
\label{fig18}
\end{figure}

First, let us show the results for $V_a=0.4$, $V_q=0.7,$ $V_{2c}=1.1$.
In this case, we find that the spin singlet pairing dominates over 
triplet pairing.
The obtained spin-singlet gap function is shown in Fig.\ref{fig15} 
$(\Delta_{\rm singlet})$ 
along with the Fermi surface. The gap function changes sign as 
$+-+-$ along the Fermi surface with nodes intersecting 
at positions on the $k_x$ and $k_y$ 
axes as in the $d_{xy}$ pairing. However, the entire gap function 
roughly has the form 
\begin{eqnarray}
\Delta(\Vec{k})&\propto& \cos(k_x-k_y)-0.8\cos(k_x+k_y)\nonumber\\
&+&0.3[\cos(3k_x+2k_y)+\cos(2k_x+3k_y)]\nonumber\\
&-&0.2[\cos(2k_x-k_y)+\cos(k_x-2k_y)]\nonumber\\
&-&0.2\cos(4k_x+4k_y),
\end{eqnarray}
so that there is a deviation from the simple $d_{xy}$ gap form 
$\cos(k_x-k_y)-\cos(k_x+k_y)\propto \sin(k_x)\sin(k_y)$.
The additional terms such as $\cos(3k_x+2k_y)$, 
$\cos(2k_x-k_y)$, and $\cos(4k_x+4k_y)$ 
imply that the pairing occurs to some extent 
at distances $(\Delta x,\Delta y)=$ 
$\pm(3,2)$,$\pm(2,-1)$, and $\pm(4,4)$ 
(in units of the lattice constants in the $x$ and $y$ directions), 
respectively, which can be considered as a consequence of the distant off-site 
repulsions that push away the electrons to far distances.

$T_c$ is plotted as a function of $t_c$ in Fig.\ref{fig16}(a), where 
it is enhanced in the vicinity of the charge ordering phase. 
For comparison, 
we also plot $T_c$ in the absence of all the off-site repulsions, where 
the pairing is solely due to the spin fluctuations, and 
the charge fluctuations virtually do not contribute.
Since $T_c$ is higher in the presence of the off-site repulsions while 
the pairing symmetry is unchanged,  
we can say that 
the charge fluctuations cooperate with the spin fluctuations to 
enhance this superconductivity. This is also seen in Fig.\ref{fig17}(a), 
where we plot the maximum eigenvalue of the gap equation $\lambda$ 
for the singlet and the triplet pairings for cases in which all the 
off-site repulsions $(V_c,V_p,V_a,V_q,V_{2c})$ 
are multiplied by a factor $\beta$ ($\beta=1$ corresponds to 
$V_c=2$, $V_p=1.8$, $V_a=0.4$, $V_q=0.7$, $V_{2c}=1.1$).
Here we also plot a ratio $\gamma$ defined as 
\begin{equation}
\gamma=\frac{[-\{U+2V(\Vec{q})\}\chi_0(\Vec{q})]_{\rm max}}
{[U\chi_0(\Vec{q})]_{\rm max}}.
\end{equation}
The denominator and the numerator in this ratio are terms that appear in the 
denominator of the spin and charge susceptibilities, respectively, 
and the subscript ``max'' denotes the maximum value within the entire BZ.
Roughly speaking,  
the charge fluctuations dominate over 
spin fluctuations (the peak value of the charge susceptibility 
exceeds that of the spin susceptibility) when $\gamma$ exceeds unity. 
We can see from Fig.\ref{fig17}(a) 
that in the charge fluctuation dominating 
regime, the spin singlet pairing (as well as the triplet pairing) is 
indeed enhanced as the off-site repulsions are increased. 

To see this point in more detail, we plot 
the spin and charge susceptibilities, the singlet and triplet 
pairing interactions and the gap functions in Fig.\ref{fig15}. 
The spin susceptibility has a peak around wave vector (1),
which bridges the portions of the Fermi surface 
with the same sign of the gap in the singlet $d_{xy}$ pairing, and therefore
work destructively against this pairing. 
(Note that 
$\Delta(\Vec{k})\Delta(\Vec{k'})V^{pair}(\Vec{k}-\Vec{k'})<0$, 
$\Vec{k},\Vec{k}'\in {\rm Fermi\:\: surface}$,  
is favorable for superconductivity with a gap function 
$\Delta(\Vec{k})$.)
However, the 
spin susceptibility takes relatively large values 
also around wave vectors (3)$\sim$(2), 
which bridges the 
portions of the Fermi surface with different gap signs in the $d_{xy}$
pairing. So the spin fluctuations have components 
which favor (around (1)) and 
unfavor (around $(2)\sim (3)$) 
the $d_{xy}$ pairing. On the other hand, the charge 
susceptibility strongly peaks near position (1), very close to the 
peak position of the spin susceptibility, at which the spin fluctuations 
work destructively against the $d_{xy}$ pairing. The charge fluctuation 
around position (1) is so strong that it cancels out the spin fluctuation 
contribution (see eq.(\ref{pairsinglet})) 
and even turns the singlet pairing interaction ($V_{singlet}^{pair}$ in 
Fig.\ref{fig15}) into a negative value at this position, thereby 
enhancing the $d_{xy}$ pairing. 
%

Now let us turn to the case of $V_a=V_q=0.6$, $V_{2c}=1.1$. 
Although the changes in the parameter values are not so large, and the 
peak position of the charge susceptibility for $t_c=-0.3$ is very 
close to that in the case of $V_a=0.4$, $V_q=0.7$, $V_{2c}=1.1$, 
we find that spin triplet pairing with a gap shown in 
Fig.\ref{fig18} ($\Delta_{\rm triplet}$) dominates over singlet in this case.
$T_c$ of the triplet pairing is shown as a function of $t_c$ in 
Fig.\ref{fig16}(b), where superconductivity is again enhanced in 
the vicinity of the charge ordering phase.
The reason why spin-triplet pairing is enhanced is because the charge 
fluctuations arise around positions $(1)\sim (2)$ 
(note that $(\frac{1}{2},k_y)$ and $(-\frac{1}{2},k_y)$ 
are identical points) shown in  
Fig.\ref{fig18}, thus giving rise to 
attractive triplet pairing interaction around position (2),
which favors the gap $\Delta_{\rm triplet}$ because vector (2) 
bridges the portions of the Fermi surface that have the same gap sign.
An analysis similar to 
the one done for the singlet pairing shows that the 
gap roughly has the form 
\begin{eqnarray}
\Delta(\Vec{k})&\propto& \sin(k_x+2k_y)-\sin(2k_x+k_y)\nonumber\\
&-&0.3\sin(k_x-k_y)
+0.3[\sin(2k_x)-\sin(2k_y)]\nonumber\\
&-&0.3[\sin(4k_x)-\sin(4k_y)]\nonumber\\
&+&0.3[\sin(k_x-2k_y)-\sin(k_y-2k_x)].
\end{eqnarray}
The dominating part has the form 
$\propto \sin(k_x+2k_y)-\sin(2k_x+k_y)$, 
so we will call this pairing $p_{x+2y}$-wave pairing.
Although the pairing occurs mainly between sites separated by 
$\pm (1,2)$ and $\pm (2,1)$,  
here again, the pairs are also formed, to some extent, at far distances.
The difference from the case of $V_a=0.4$, $V_q=0.7$ is 
that the charge fluctuations tend to spread out toward position (2) 
in the case of $V_a=V_q=0.6$, 
so that the absolute value of the 
attractive triplet pairing interaction around position (2)  
is larger 
($\sim 50$ for $V_a=V_q=0.6$ and $\sim 20$ for $V_a=0.4$, $V_q=0.7$ ).
\cite{comment2} 
In the case of $V_a=0.4$, $V_q=0.7$, 
the charge fluctuations are more localized around position (1), at 
which the charge fluctuations do 
not contribute to $p_{x+2y}$-wave pairing because 
vector (1) bridges the portion of the Fermi surface at which 
the nodes of the $p_{x+2y}$-wave gap exist. This difference of the 
charge susceptibility between the two cases 
originates from the structure of $V(\Vec{q})$ 
discussed in section \ref{rpares}  (note that the value of 
$t_c$ does not affect $V(\Vec{q})$), namely, $-V(\Vec{q})$ is larger 
around $\Vec{q}=(\frac{1}{2},-\frac{1}{6})_{uf}
\sim (\frac{11}{24},-\frac{5}{24})_{uf}$, 
i.e., position (2), for smaller $V_q$ and 
larger $V_a$, so that the charge fluctuations spread out toward position 
(2) even when $t_c>0$, although the peak itself is at position (1) for 
$t_c>0$ because the nesting vector is now close to that position.
As a consequence of the difference in the charge fluctuations,
a difference arises also in the spin singlet pairing gap function. 
Namely, the strong charge fluctuations around position 
(2) in the case of $V_a=V_q=0.6$ 
results in attractive 
singlet pairing interaction not only near position (1) as 
in $V_a=0.4$, $V_q=0.7$, but also near position (2), 
which works destructively against $d_{xy}$ pairing, resulting in 
a vanishing gap around $k_x=-k_y$ (end point of vector (2) in 
$\Delta_{\rm singlet}$ of Fig.\ref{fig18}) in the  
singlet channel, thereby giving way to spin-triplet pairing.
In fact, we can see from Fig.\ref{fig17} that triplet pairing dominates 
over singlet in the charge fluctuation dominating regime when 
$(V_a,V_q)=\beta(0.6,0.6)$ (Fig.\ref{fig17}(b)), while 
singlet pairing continues to dominate deep within the charge fluctuation 
dominating regime when $(V_a,V_q)=\beta(0.4,0.7)$ (Fig.\ref{fig17}(a)).

So the bottom line here for the superconductivity in the 
case of $t_c>0$ is that spin triplet pairing 
tends to dominate over spin singlet pairing
when the off-site interaction values are those that  give 
strong charge fluctuations (namely high $T_{\rm co}$) for $t_c<0$ 
near the position where the diffuse X-ray spots are 
observed in $X=MM'$(SCN)$_4$. 
This tendency is more systematically 
seen in Fig.\ref{fig10}, where we classify the values of 
$(V_a,V_q)$ at which triplet (singlet) pairing dominates for $t_c=+0.2$ 
by red (green) circles. 
Considering the ambiguity in the 
values of the on-site and off-site repulsions, we cannot 
predict at the present stage 
whether singlet or triplet pairings dominate in $X=$I$_3$.
The singlet-triplet competition is subtle, but in any case, 
the presence of the charge fluctuations enhances  
superconductivity.
%
%
%

\section{Comparison with (TMTSF)$_2$X}

The coexistence of spin and charge fluctuations near 
the nesting vector ``$\Vec{Q}_{2k_F}$'' of the Fermi surface 
(although the nesting is not 
so good for $t_c>0$) is reminiscent 
of a quasi-one-dimensional organic superconductor 
(TMTSF)$_2$PF$_6$, in which we have proposed a pairing mechanism 
due to the coexistence of $2k_F$ spin and $2k_F$ charge fluctuations.
\cite{KAA,YK,KY,Kuroki} In this material, coexistence of 
$2k_F$ SDW and $2k_F$ CDW has been experimentally observed in the 
insulating phase sitting next to the superconducting phase in the 
pressure-temperature phase diagram.\cite{PR,Kagoshima}
This  coexistence cannot be understood within a 
model that considers only  the on-site $U$ and the 
nearest neighbor repulsion $V$, and the 
effect of the second nearest neighbor repulsion $V'$ is crucial
\cite{KobayashiOgata,Suzumura,YK,KY,Fuseya,Nickel}, 
sharing similarity with the present case. 
In fact, the mechanism of $2k_F(=\frac{\pi}{4}\times2=\frac{\pi}{2})$ 
CDW in TMTSF is very similar to 
the mechanism of $3\times (3\sim 4)$ ordering proposed in the 
present study. Namely, if we consider only the nearest ($V$) and the next 
nearest neighbor ($V'$) interactions within the 
chains, the Fourier transform of the off-site repulsions 
is given as $V(\Vec{q})=2V\cos(q_x)+2V'\cos(2q_x)$. On the other hand, 
the bare susceptibility $\chi_0(\Vec{q})$ has a peak 
at $q_x=\frac{\pi}{4}$ due to a good nesting of the Fermi surface of the  
quarter-filled quasi-1D band. Since both $-V(\Vec{q})$ and 
$\chi_0(\Vec{q})$ take large positive values at $q_x=\frac{\pi}{2}$,
$2V(\Vec{q})\chi_0(\Vec{q})$ 
in the denominator of eq.(\ref{chargeRPA}) takes a large negative 
value there, resulting in a peak in the charge 
susceptibility at $2k_F$. 
Therefore, within this scenario, the $2k_F$ CDW in 
(TMTSF)$_2$PF$_6$ occurs precisely due to a 
cooperation between the effect of the off-site (next 
nearest neighbor) repulsions and the Fermi surface nesting.

In the case of (TMTSF)$_2$PF$_6$, 
due to the good nesting of the Fermi surface and the one dimensionality, 
the $2k_F$ peak positions as well as the overall $\Vec{q}$ dependence of 
the spin and charge susceptibilities nearly coincide, 
so that the two fluctuations work destructively 
in the singlet pairing (eq.(\ref{pairsinglet}))
while constructively in the triplet pairing (eq.(\ref{pairtriplet})). 
Also, due to the quasi-one-dimensionality of the system, the 
Fermi surface is disconnected in a manner that 
spin singlet $d$-wave-like pairing and spin-triplet $f$-wave-like 
pairing have the same number of nodes on the Fermi surface.\cite{KAA}
Due to these factors favoring spin triplet pairing, 
charge fluctuations with similar strength with the 
spin fluctuation result in a subtle competition between 
$f$-wave and $d$-wave pairings in a realistic parameter 
regime.\cite{KY,Kuroki,Nickel}

In the present case of the $\theta$-(BEDT-TTF) compounds,
there are several differences as compared to TMTSF:
(i) there is a large component in the spin fluctuations 
that work destructively against singlet pairing, 
(ii) the charge fluctuations cancel out this spin fluctuation component 
in the singlet pairing interaction to enhance the pairing,
(iii) The singlet pairing has smaller number of 
nodes on the Fermi surface than in the triplet 
pairing because there is no disconnectivity in the Fermi surface. 
(iv) since there are 6 nearest neighbor 
($V_p$, $V_c$) and 8 next nearest neighbor ($V_a$, $V_q$, $V_{2c}$) 
interactions in the $\theta$ compounds as compared to 4 nearest 
 (including the interchain direction) and 2 next nearest neighbor interactions 
in (TMTSF)$_2$X,  $V(\Vec{q})$ tends to be large, so that 
the charge fluctuations are likely to strongly dominate over spin fluctuations 
in a realistic parameter regime. In the charge fluctuation dominating regime, 
the charge fluctuations enhance 
both singlet and triplet pairings, but the enhancement in the latter 
is stronger.
Factors (i) and (iv) favor triplet pairing, while (ii) and (iii) singlet 
pairing. Due to these competing factors, there is a close competition 
between singlet and triplet pairing once again as in (TMTSF)$_2$X.

\section{Conclusion} 
In the present study, we have investigated the origin of the 
charge orderings in $\theta$-(BEDT-TTF)$_2MM'$(SCN)$_4$. We have shown that 
neither the low temperature horizontal stripe nor the 
high temperature $3\times (3\sim 4)$ charge orderings  
can be understood within a model that takes into account only 
$U$, $V_p$, and $V_c$ as electron-electron repulsions.
We have shown that the 
$3\times (3\sim 4)$ charge ordering 
can be considered as a consequence of the cooperation between the 
effect of the off-site repulsions including the 
distant interactions $V_{a}$, $V_{q}$, and $V_{2c}$,  
and the nesting of the Fermi surface.
Moreover, we have proposed a possibility that the 
horizontal stripe charge ordering is triggered by the 
new $(0,\frac{1}{2})_f$ nesting in the $3\times (3\sim 4)$  charge-ordered 
state, where some portions of the Fermi surface remain ungapped.
From this viewpoint, we can understand why the 
horizontal stripe charge ordering temperature
decreases with increasing $t_c$ (from $t_c<0$ to $t_c>0$) in 
$\theta$-(BEDT-TTF)$_2$X. Namely, as $t_c$ increases, the 
nesting of the Fermi surface is degraded, the maximum positions 
of $\chi_0(\Vec{q})$ and $-V(\Vec{q})$ deviate, and the density of 
states becomes small, all working destructively against 
$3\times (3\sim 4)$ ordering. Since $3\times (3\sim 4)$ 
``first stage'' ordering gives 
base for the horizontal stripe ``second stage'' 
ordering in our view, the temperature of the 
latter ordering should also become low as $t_c$ increases.

In the present study, we have obtained the Fermi surface of the $3\times 3$ 
charge ordered state by 
assuming that this first stage ordering 
is truly long ranged, while it is not a true long range order in the actual 
materials. Also, the role of the electron-electron interactions in 
the occurrence of the second stage ordering as well as the 
second stage ordering temperature 
has not been addressed quantitatively. 
A more detailed and quantitative understanding of this 
``successive charge ordering'' view, as well as the investigation on its 
possible relevance to the nonlinear transport,\cite{Inagaki,Sawano}
 serves as an interesting future study.

Given that the distant interactions are important in 
$\theta$-(BEDT-TTF)$_2$X, we have further investigated the 
possibility of superconductivity in the vicinity of the 
charge ordering phase for $t_c>0$. We have shown that there is a 
close competition between  $d_{xy}$-wave-like 
singlet pairing and $p_{x+2y}$-wave-like triplet pairing, 
where the latter tends to dominate over the former when we 
adopt the interaction values that give strong charge fluctuations 
(high $T_{\rm co}$) around $(\frac{2}{3},\frac{1}{3}\sim\frac{1}{4})_f$ 
in the case of $t_c<0$.
Regardless of whether triplet pairing dominates or not, 
the charge fluctuations are found to enhance 
superconductivity.

In total, we have provided a unified understanding of the overall 
phase diagram of $\theta$-(BEDT-TTF)$_2$X\cite{HMori2} 
at least qualitatively within an electronic model that considers 
the band structure and the electron-electron interactions 
up to next nearest neighbors.

\acknowledgement
The author would like to thank  H. Mori, T. Mori, T. Yamaguchi, H. Seo, 
and R. Arita for discussions.
The author acknowledges 
Grants-in-Aid for Scientific Research from the Ministry of Education, 
Culture, Sports, Science and Technology of Japan, and from the Japan 
Society for the Promotion of Science.
Part of the numerical calculation has been performed
at the facilities of the Supercomputer Center,
Institute for Solid State Physics, University of Tokyo.
\ \\

%
%
%


\begin{thebibliography}{99}
\bibitem{comment} The reason we chose values of interactions to be slightly 
larger than that in the RPA calculation is because $\Delta n_{\rm max}$, if 
any, becomes very small if we adopt the values used in the RPA. 
This is probably because we assume 
no ordering in the spin density in the present mean field calculation, 
whereas in the RPA calculation, the spin susceptibility 
does have a well defined peak at the nesting vector.
\bibitem{comment4} Due to symmetry, the opposite can also happen with the 
same probability, i.e., the $3\times 3$ ordering using the 
$(-\frac{2}{3},\frac{1}{3}\sim \frac{1}{4})_f$ nesting, resulting in a 
new nesting near $(0,\frac{1}{2})_f$  
due to the original $(+\frac{2}{3},\frac{1}{3}\sim \frac{1}{4})_f$ 
nesting. There is also 
a possibility that the two charge orderings with different 
modulation simultaneously set in at the same 
temperature, resulting in 
a state with an almost fully gapped Fermi surface. 
Although in a strict sense 
this possibility has to be addressed theoretically,  
such a possibility is unlikely from an experimental point of view  
considering that the system remains metallic and no X-ray spot is 
observed at $(0,\frac{1}{2})_f$ 
in the $3\times (3\sim 4)$ ordering state at high temperatures.
\bibitem{comment3}
If the $a$ component of vector (2) deviates from $k_a=-\frac{2}{3}$,
the $a$ component of vector (3) deviates from $k_a=0$. In that case, 
if we consider the ``opposite case'' mentioned in comment \citen{comment4}, 
the new nesting vector appears at a position  
symmetric with respect to the $k_c$ axis. In the actual experiments, the  
mixture of these two opposite cases should be observed.
\bibitem{comment2} To be more precise, the peak absolute value of the 
triplet pairing interaction at position (1) itself is also larger 
for $V_a=V_q=0.6$ and $t_c=0.2$ $(\sim 90)$ than for 
$V_a=0.4, V_q=0.7$ and $t_c=0.14$ $(\sim 50)$. 
However, if we take 
$t_c=0.12$ for $V_a=0.4, V_q=0.7$, the peak absolute value at position (1)
($\sim 160$) by far exceeds that for 
$V_a=V_q=0.6$ and $t_c=0.2$, 
but still the absolute value at position (2) 
remains to be $\sim 20$, meaning that the peak structure 
is strongly localized. 
\end{thebibliography}
\end{document}